\documentclass[superscriptaddress]{article}
\usepackage{amssymb,amsthm,amsmath,amsfonts,physics,algorithm2e,float}
\usepackage[a4paper, total={7in, 9in}]{geometry}
\usepackage{soul}
\usepackage{graphicx}
\usepackage[pdftex,dvipsnames,usenames]{xcolor}
\usepackage{authblk}
\usepackage[colorlinks=true,urlcolor=blue,citecolor=blue,linkcolor=blue]{hyperref}
\usepackage[sorting=none,backend=biber,style=nature]{biblatex}
\usepackage{multicol}
\usepackage{caption}
\usepackage{subcaption}
\usepackage[dvipsnames]{xcolor}
\usepackage{siunitx}
\title{Quantum teleportation with dissimilar quantum dots over a hybrid quantum network}
\author[1]{Alessandro Laneve\thanks{alessandro.laneve@uniroma1.it}}
\author[1]{Giuseppe~Ronco}
\author[1]{Mattia~Beccaceci}
\author[1]{Paolo~Barigelli}
\author[2]{Francesco~Salusti}
\author[2]{Nicolas~Claro-Rodriguez}
\author[1]{Giorgio~De Pascalis}
\author[1]{Alessia~Suprano}
\author[1]{Leone~Chiaudano}
\author[2]{Eva Schöll\thanks{current affiliation: Institute of Semiconductor and Solid State Physics, Johannes Kepler University, Altenbergerstraße 69, Linz 4040, Austria}}
\author[2]{Lukas Hanschke\thanks{current affiliation: Walter Schottky Institut, TUM School of Computation, Information and Technology and MCQST, Technical University of Munich, Hans-Piloty Str. 1, 85748 Garching, Germany}}
\author[3]{Tobias~M.~Krieger}
\author[4]{Quirin~Buchinger}
\author[3,5]{Saimon~F.~Covre~da~Silva}
\author[1]{Julia~Neuwirth}
\author[6]{Sandra~Stroj}
\author[4]{Sven~Höfling}
\author[4]{Tobias~Huber-Loyola}
\author[7]{Mario~A.~Usuga Castaneda}
\author[1]{Gonzalo~Carvacho}
\author[1]{Nicolò~Spagnolo}
\author[1]{Michele~B.~Rota}
\author[1]{Francesco~Basso Basset}
\author[3]{Armando~Rastelli}
\author[1]{Fabio~Sciarrino}
\author[2]{Klaus~Jöns}
\author[1]{Rinaldo~Trotta\thanks{rinaldo.trotta@uniroma1.it}}

\affil[1]{Dipartimento di Fisica, Sapienza Università di Roma, Piazzale Aldo Moro 5, I-00185 Roma, Italy}
\affil[2]{Institute for Photonic Quantum Systems (PhoQS), Center for Optoelectronics and Photonics Paderborn
(CeOPP) and Department of Physics, Paderborn University, Warburger Straße 100, 33098, Paderborn, Germany}
\affil[3]{Institute of Semiconductor and Solid State Physics, Johannes Kepler University, Altenbergerstraße 69, Linz 4040, Austria}
\affil[4]{Technische Physik, University of Würzburg, Am Hubland, D-97074 Würzburg, Germany}
\affil[5]{Universidade Estadual de Campinas, Instituto de F\'{i}sica Gleb Wataghin, 13083-859 Campinas, Brazil}
\affil[6]{Forschungszentrum
Mikrotechnik, FH Vorarlberg, Hochschulstr. 1, A-6850 Dornbirn, Austria}
\affil[7]{Single Quantum B.V., Delft, HH 2629, The Netherlands}

\begin{document}

\maketitle
\begin{abstract}
\textbf{Photonic quantum information processing in metropolitan quantum networks lays the foundation for cloud quantum computing \cite{barz2012demonstration,maring2024versatile}, secure communication \cite{yin2020entanglement,paraiso2021photonic}, and the realization of a global quantum internet \cite{kimble2008quantum,wehner2018quantum}.  This paradigm shift requires on-demand and high-rate generation of flying qubits and their quantum state teleportation over long distances \cite{ren2017ground}. Despite the last decade has witnessed an impressive progress in the performances of deterministic photon sources \cite{mckeever2004deterministic,somaschi2016near,ding2016demand,tomm2021bright}, the exploitation of distinct quantum emitters to implement all-photonic quantum teleportation among distant parties has remained elusive. Here, we overcome this challenge by using dissimilar quantum dots whose electronic and optical properties are engineered by light-matter interaction \cite{liu2019solid}, multi-axial strain \cite{huber2018strain} and magnetic fields \cite{bayer1998exciton} so as to make them suitable for the teleportation of polarization qubits. This is demonstrated in a hybrid quantum network harnessing both fiber connections and 270 m free-space optical link connecting two buildings of the University campus in the center of Rome. The protocol exploits GPS-assisted synchronization, ultra-fast single photon detectors as well as stabilization systems that compensate for atmospheric turbulence. The achieved teleportation state fidelity reaches up to $82\pm1\%$, above the classical limit by more than 10 standard deviations. Our field demonstration of all-photonic quantum teleportation opens a new route to implement solid-state based quantum relays and builds the foundation for practical quantum networks.
 }
\end{abstract}

\begin{multicols}{2}
\begin{figure*}[t!]
    \centering
\includegraphics[width=\linewidth]{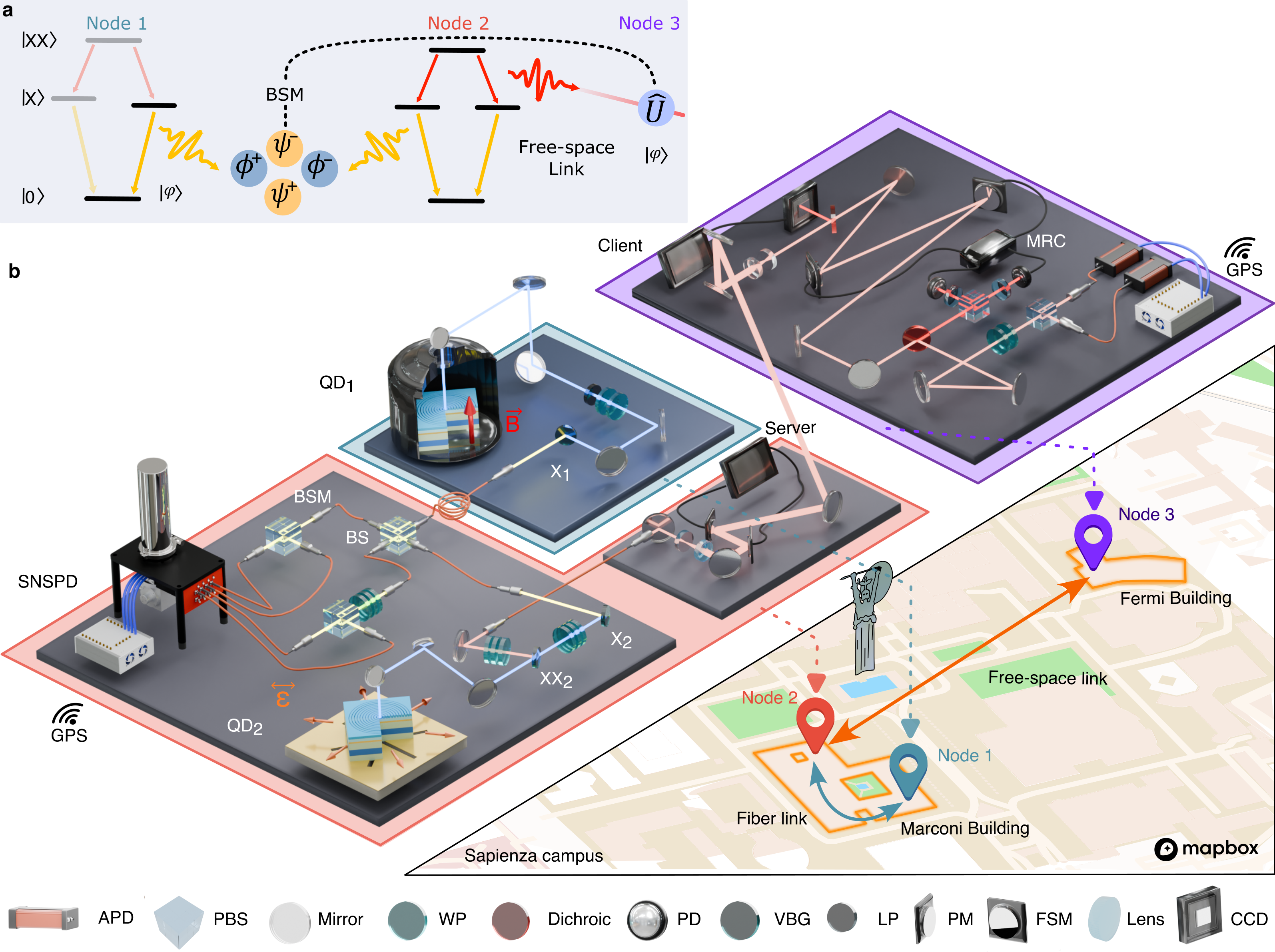}
    \caption{\textbf{Sketch of the urban quantum teleportation network.} 
      \textit{\textbf{a}, Scheme of the quantum teleportation protocol as implemented with two QDs over a network including a free-space link. \textbf{b}, Depiction of the teleportation experimental realization over the Sapienza University campus. In Node 1, highlighted in teal and placed in the Marconi building, the X$_1$ photon from QD$_1$ is selected and prepared in a well-defined polarization state.
      Node 1 is connected with a 15 m optical fibre link to Node 2, highlighted in red,
      where QD$_2$ emits a polarization-entangled biexciton-exciton photon pair, XX$_2$ and X$_2$. Spatial displacement between Node 1 and Node 2 is accentuated on the map for clarity. A BSM is performed in Node 2, while XX$_2$ is sent to Node 3, highlighted in purple and placed in the Fermi building through a 270 m free-space optical link. The strain and the magnetic field harnessed to optimize the X$_2$-XX$_2$ degree of entanglement and the X$_1$ emission wavelength, are featured in orange and red, respectively. The inset a map of the Sapienza campus, where the Marconi and Fermi buildings are framed in orange. The experimental setup accommodates Volume-Bragg Gratings (VBG), polarizing beam-splitters (PBS), waveplates (WP), piezoelectric mirrors (PM), fast steering mirrors (FSM), and client-server PCs for the optimization of the free-space optical link.}}
    \label{fig:1-sketch_network} 
\end{figure*}
Large scale quantum networks require the capability of transferring quantum information between distant nodes. Currently, photons represent the best candidate to carry out such a task: they are compatible with pre-existing optical fiber \cite{mao2018integrating} and free-space infrastructure \cite{yin2020entanglement}, they are extremely resilient to environmental decoherence and they feature a variety of easily manipulable degrees of freedom, for which quantum correlations have been demonstrated \cite{kwiat1997hyper,mair2001entanglement,simon2005creating, tanzilli2005photonic}. More specifically, the
flexibility and noise resilience of the polarization
degree of freedom makes it the current optimal solution to
distribute quantum states and quantum correlations
over long distances \cite{yin2017satellite, ren2017ground}. 
 However, photon distribution over the globe will inevitably be affected by losses, stressing the need of \textit{quantum relays}, i.e. "devices" that mitigate overall losses by teleporting quantum information between distant nodes of a quantum network \cite{gisin2007quantum}. 
The all-photonic version of quantum relays inherently sets very stringent requirements on the sources of indistinguishable and
entangled photon states employed across the network. These include high production rate, wavelength tunability, the need of high photon indistinguishability, and near-unity degree of entanglement, to mention a few.
 In this perspective, semiconductor quantum dots (QDs) have been under the spotlight for two decades, mainly due to the lack of trade-off between brightness and single photon purity (or degree of
entanglement) that is typically plaguing Poissonian
sources of light \cite{lodahl2017quantum}. 
More specifically, QDs have shown on-demand generation of single and entangled photons with high
indistinguishability \cite{scholl2019resonance,zhai2022quantum}, near-unity degree of
entanglement \cite{huber2017highly}, ultra-low multi-photon emission
\cite{schweickert2018demand} and, most importantly, high brightness \cite{dousse2010ultrabright,somaschi2016near,liu2019solid}.
Yet, the use of independent QDs to implement all-photonic quantum teleportation in a quantum relay scenario has evaded demonstration so far. The main reason is that this task demands state-of-the-art single and entangled photon sources to be interfaced via two-photon interference.
This, in turn, requires the indistinguishability of photons generated by independent QDs, which inevitably feature dissimilar optical properties. Despite pioneering steps along this direction have been moved \cite{gao2013quantum,schimpf2021quantum,zhai2022quantum}, the requirements set by all-photonic teleportation protocols are simply too demanding to be met by conventional technologies or cherry-picking approaches. To better explain this hurdle, it is sufficient to consider that the "simple" task of generating single photons with high efficiency and indistinghuishability requires the use of sophisticated nanophotonic devices \cite{santori2002indistinguishable,ates2009post,gazzano2013bright} that exploit light-matter interaction to boost the flux of QD photons.
The level of complexity increases dramatically if QDs are to deliver entangled photons with high efficiency, mainly because their electronic structure has to be controlled with high accuracy \cite{bayer2002fine} and the generated photon pairs feature dissimilar energies. This usually requires additional manipulation techniques in specifically designed optical cavities \cite{dousse2010ultrabright,gregersen2016broadband,liu2019solid}. Quantum relays with remote QDs not only require all these provisions to be met, but add an additional twist: the need to interface dissimilar and distant single and entangled photon sources. Considering the efforts required to just gather the necessary ingredients, putting them together may be fairly regarded as a titanic endeavor.\\ 
Here, we harness two decades of research in QD quantum photonics to show that this challenge can be overcome. We first fabricate state-of-the-art QDs embedded in Circular Bragg Resonators (CBR) \cite{liu2019solid} designed to boost the flux of both single and entangled photon pairs. We then use external perturbations, specifically multi-axial strain and magnetic fields, to reshape the electronic structure of dissimilar QDs so as to make them suitable for quantum teleportation. Finally, we take advantage of ultrafast nanowire single photon detectors to mitigate the effects of residual photon distinghuishability. These ingredients enable the first demonstration of all-photonic quantum state teleportation with dissimilar QDs in an urban communication scenario, specifically in a hybrid quantum network laid over the Sapienza University campus in Rome.\\
Our achievement is qualitatively showcased in Fig. \ref{fig:1-sketch_network}. Two independent and remote QDs are driven under two-photon resonant excitation (TPE) \cite{jayakumar2013deterministic} to generate nearly on-demand photon pairs through the biexciton-exciton radiative cascade \cite{muller2014demand}. We denote the photon emitted by the exciton recombination as X, and the one after the biexciton recombination as XX (Fig. \ref{fig:1-sketch_network}\textbf{a}).
The input state $\ket{\phi}$ is prepared in the polarization degree of freedom of the X$_1$ photons, generated by QD$_1$ in a first laboratory, acting as Node 1 (see Fig. \ref{fig:1-sketch_network}\textbf{b}). 
This photon is sent to a second laboratory, Node 2, where QD$_2$ generates the XX$_2$-X$_2$ entangled photon pair. There, the quantum interference between X$_1$ and X$_2$ photons enables the teleportation of the $\ket{\phi}$ target state onto the polarization state of XX$_2$, which is entangled to X$_2$. We first benchmark the teleportation between the two separated nodes 1 and 2 in the Marconi building of the Sapienza Physics Department, realizing the first demonstration of all-photonic quantum teleportation between distinct QDs. Then, the XX$_2$ photon is sent via a 270 m free-space optical link to a third laboratory in a different building of the campus (named after Fermi), acting as Node 3.
The link is equipped with specifically designed synchronization devices, and stabilization systems to compensate for atmospheric turbulence.
Overall, we achieve the successful teleportation of a polarization qubit using dissimilar QDs in an urban quantum network comprising both fiber and free-space links.
The key to the success of the experiments lies in the use of state-of-the-art photonic cavities in combination with innovative quantum-engineering techniques (as sketched in Fig. \ref{fig:1-sketch_network}\textbf{b}) to make QD$_1$ and QD$_2$ suitable for quantum teleportation, as explained in the next section.



\section*{Engineering quantum light sources}\label{sec2}
\begin{figure*}[t!]
    \centering    \includegraphics[width=\linewidth]{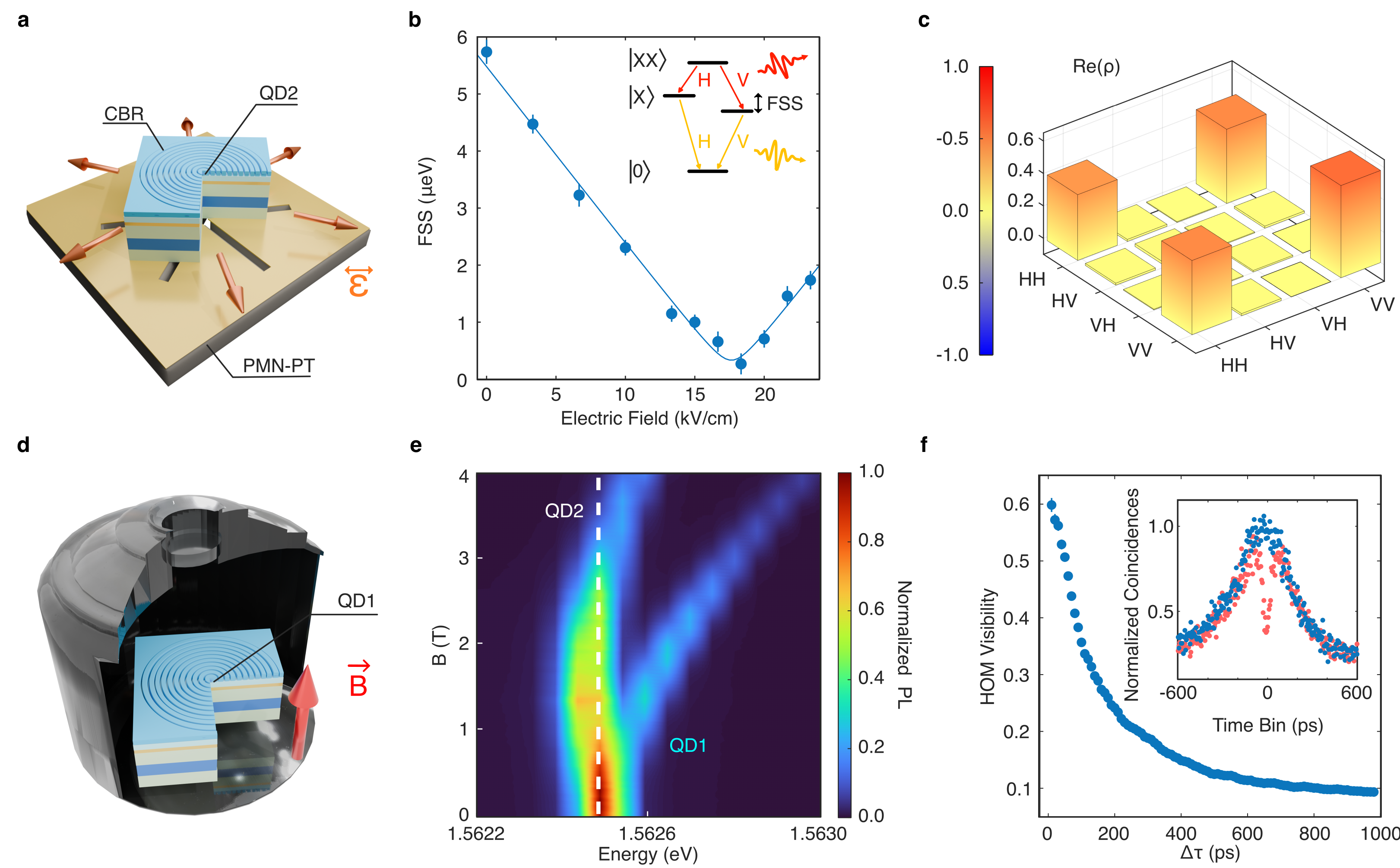}
    \caption{\textbf{Quantum resources engineering.} \textit{\textbf{a}, Schematic representation of the CBR QD integrated onto PMN-PT to enable the application of mechanical stress. \textbf{b}, The strain-tuning curve along one axis showing our capability to erase the FSS. The solid line is a fit from the model equations reported in \cite{trotta2012universal}. 
    \textbf{c}, Experimental two-photon density matrix (real part) measured for FSS$=0.3\pm0.2$ $\mu$eV, yielding a fidelity $F_{\ket{\phi^+}}=0.94\pm0.01$ and a concurrence $C=0.89\pm0.01$. 
    \textbf{d} Schematic representation of QD$_1$ inserted in a magneto-cryostat.
    \textbf{e}, Diamagnetic shift and Zeeman splitting of the emission spectrum of QD$_1$: thanks to a fine tuning of the $B$ intensity, we achieve spectral overlap between the two X photons applying a magnetic field of $B=0.9$ T.
    \textbf{f}, HOM visibility between photons from remote QDs as a function of the coincidence time window $\Delta \tau$: post-selection on the HOM events allows us to reach a maximum HOM visibility of $V_{HOM}=0.598\pm0.025$ for a $20$ ps time window.
    Inset: time distribution of coincidences in the HOM experiment for co-polarized photon pairs (red) and cross-polarized photon pairs (blue). 
    }}
    \label{fig:2-source_characterization} 
\end{figure*}
\begin{figure*}[t!]
    \centering
\includegraphics[width=\linewidth]{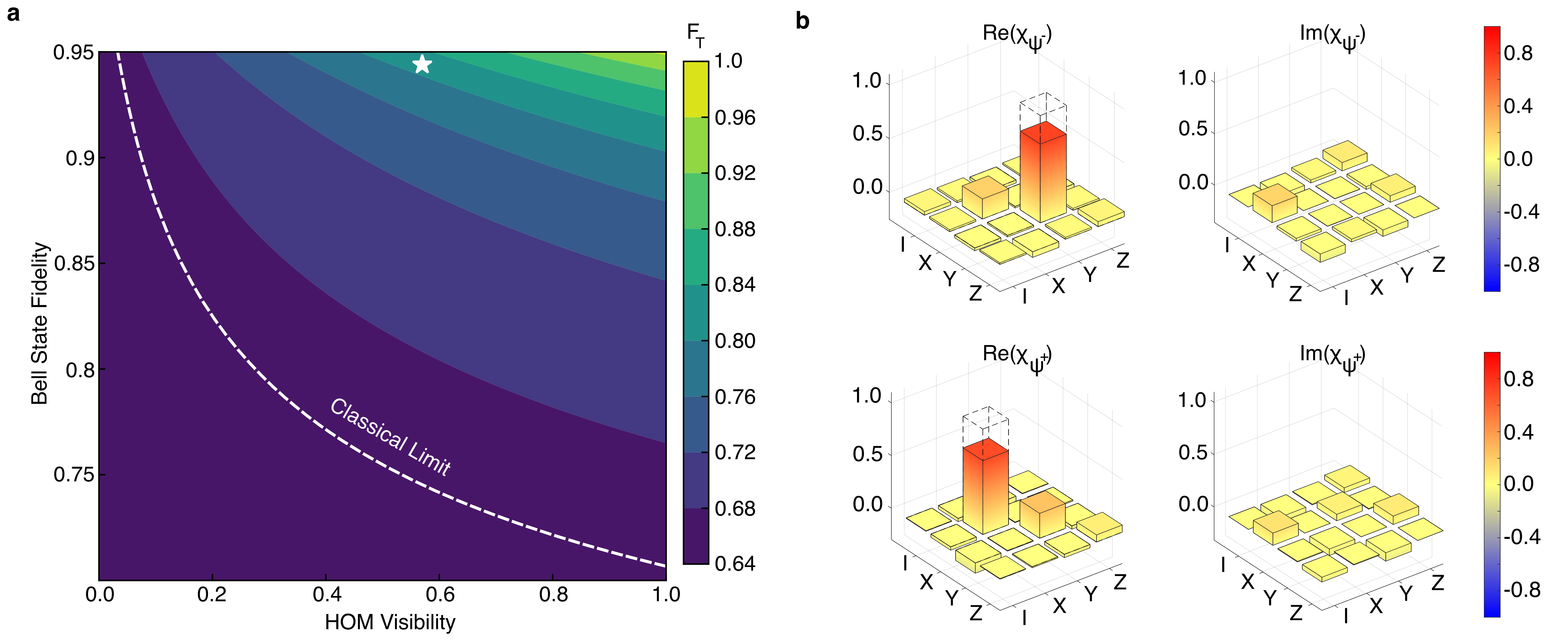}
    \caption{\textbf{Quantum state teleportation.}  \textit{ \textbf{a}, Theoretical teleportation fidelity as a function of the quantum resources available for the protocol, computed for the case of a $50\%$ BSM. The white curve highlights the maximum achievable fidelity with classical resources, i.e. for a protocol only relying on classical correlations. The white star points out the expected teleportation fidelity as estimated with our theoretical model. In particular, our Bell state fidelity is $0.94\pm0.01$ and the HOM visibility for $30$ ps coincidence window is $V_{HOM}=0.57\pm0.02$, that yield an average teleportaion fidelity $F_T=0.827\pm0.006$.
     \textbf{b}, Experimental process matrices $\chi_{\psi^{+/-}}$ for the teleportation protocol,  as defined in \cite{chuang1997prescription}, reported both for the projection on $\ket{\psi^-}$ (top) and $\ket{\psi^+}$ (bottom), and in comparison with the corresponding ideal process matrix (dashed). The matrices are represented in terms of the quantum operations I (the identity), X$=\sigma_X$, Y$=-i\sigma_Y$ and Z$=\sigma_Z$, where $\{\sigma_{X/Y/Z}\}$ are the Pauli matrices.}}
    \label{fig:3-teleportation_results}
\end{figure*}

For a successful all-photonic quantum teleportation with two remote QDs, three fundamental ingredients are required: \textit{(i)} a bright source of highly entangled photon pairs; \textit{(ii)} a bright source of single photons with tunable energy; \textit{(iii)} the capability to perform a successful Bell state measurement (BSM).
We accomplish \textit{(i)} by using state-of-the-art epitaxial GaAs quantum dots fabricated by the droplet etching method \cite{gurioli2019droplet}.
They can act as bright and reliable sources of single and highly polarization-entangled photon pairs, especially when embedded in suitable photonic structures \cite{liu2019solid}.
In our case, we employ GaAs QDs coupled to circular Bragg resonators (CBRs), also known as bullseye cavities, in turn integrated onto micromachined piezoelectric actuators made of [Pb(Mg$_\frac{1}{3}$Nb$_\frac{2}{3}$)O$_3$]$_{0.72}$
-[PbTiO$_3$]$_{0.28}$ (PMN-PT) \cite{rota2024source}, as depicted in Fig. \ref{fig:2-source_characterization}\textbf{a}. 
The CBR cavity engineers light-matter interaction to boost the flux of
both X and XX photons generated during the radiative cascade. Single-photon (photon-pair) extraction efficiencies as high as $85\%$ ($65\%$) have been demonstrated with this type of cavity \cite{liu2019solid}.
The micro-machined PMN-PT actuators are instead used to control the QD electronic structure via the application of independent stress fields applied along different crystal directions. These are needed to cancel out the energy splitting between the two X states - the fine structure splitting (FSS) - that is usually induced by asymmetries in the QD confining potential (see Fig. \ref{fig:2-source_characterization}\textbf{b}). Erasing this "which-path" information in the radiative cascade is fundamental to obtain near-unity degree of entanglement without temporal post-selection \cite{bayer2002fine,huber2018strain,rota2024source}.
In our experiment we employ PMN-PT in combination with CBR cavities: by varying the voltages applied to the piezoelectric actuators, we reach a  minimum value of $\text{FSS}=0.3\pm0.2$ $\mu$eV for QD$_2$, as reported in Fig. \ref{fig:2-source_characterization}\textbf{b}. By performing quantum state tomography on the generated photon pairs we estimate a fidelity of $F=0.94\pm0.01$ to the $\ket{\phi^+}$ Bell state without resorting to any time-filtering procedure (see Fig. \ref{fig:2-source_characterization}\textbf{c}), close to the best obtained with QDs in optical microcavities \cite{rota2024source}.
 The source of the teleported qubit \textit{(ii)} is another GaAs QD embedded in a CBR cavity, denoted as QD$_1$, driven under TPE and generating X$_1$ and XX$_1$ photons. The latter are not relevant for the teleportation protocol discussed here and they will not be considered further. The X$_1$ photon features a wavelength similar to X$_2$ - but not identical by several radiative linewidths. This is a clear indication that the two independent QDs feature slightly different size, shape or alloy intermixing \cite{schliwa2009few}. In order to achieve spectral indistinguishability between X$_1$ and X$_2$, we employ a magnetic field that shifts the emission by the combination of Zeeman splitting and diamagnetic shift. 
 More specifically, we select one of the Zeeman-split X$_1$ transitions and use the magnetic field to tune its energy in resonance with the X$_2$ photon, as reported in in Fig. \ref{fig:2-source_characterization}\textbf{d-e}.
 This step is crucial to achieve \textit{(iii)}. In its most efficient implementation with linear optics, a BSM exploits two photons that interfere on a symmetric beam splitter (BS) coming from two distinct input ports. Then, they undergo a polarization projective measurement at the two output ports, allowing to sample two of the four Bell states, $\ket{\psi^+}$ and $\ket{\psi^-}$.
 To perform a successful BSM, the impinging photons have to be indistinguishable in
all degrees of freedom. While the magnetic field tuning allows
us to achieve energetic resonance, X$_1$ and X$_2$ feature also
different lifetimes and linewidths, a consequence of a slightly
different Purcell effect as well as different interaction with the
solid-state environment. This can be indirectly observed in the
Hong-Ou-Mandel (HOM) interference measurement of X$_1$-X$_2$  photon pairs (realized by performing cross- and co-polarized interference, see the inset of Fig. \ref{fig:2-source_characterization}\textbf{f}), exhibiting the typical "volcano shape" \cite{legero2004quantum,scholl2020crux}.
Collecting all the photons coming out from the BS results in a limited HOM visibility, an evidence which can be fully explained by the lifetimes and linewidths of the X$_1$ and X$_2$ photons (see the SI).
However, the level of indistinghuishability can be improved by post-selecting two-photon interference events via ultrafast superconductive nanowire single photon detectors (SNSPDs) and a time-to-digital converter with 19 ps time resolution (FWHM).
By narrowing down the considered coincidence detection window $\Delta \tau$, we can achieve HOM visibilities as high as $60\%$ (see Fig. \ref{fig:2-source_characterization}\textbf{f}), a value limited by the SNSPDs time resolution. This technique clearly comes at the cost of the number of useful three-fold coincidences exploitable for the teleportation protocol, but not as much as one would intuitively expect.
In fact, we point out that the teleportation fidelity that can be achieved in an experiment depend on both the degree of entanglement and photon-indistinghuishability, and overcoming the classical limit is possible even if the available quantum resources are unbalanced, i.e., even with imperfect and dissimilar quantum dots. We have explored this possibility by using a model that computes the maximum theoretically
achievable teleportation fidelity as a function of the
degree of entanglement in the X$_2$-XX$_2$ photon pair (which is influenced by FSS and quantified by the fidelity of the photon pair state to the $\ket{\phi^+}$ Bell state) and the indistinguishaibility of the X$_1$-X$_2$ photons (measured by their HOM visibility).
The details of the model are reported in the SI while the achieved results are illustrated in Fig. \ref{fig:3-teleportation_results}\textbf{a}.
For the characteristics of our emitters, the predicted teleportation fidelity is $F_T=0.827\pm0.006$, represented by a white star in Fig. \ref{fig:3-teleportation_results}\textbf{a}.
Most importantly, the same figure highlights that the classical limit can be overcome even with a poor HOM visibility (below 20$\%$) in combination with a high level of Bell state fidelity (above 90$\%$). This suggests that successful teleportation with our independent QDs can be achieved by a loose temporal post-selection, at moderate expenses of teleportation rates. This is exactly what motivated our field demonstration of quantum teleportation, as presented in the next section.
\label{sec3}
\section*{Teleportation protocol over a quantum network}

The protocol we used builds upon the scheme employed in one of the seminal demonstration of quantum teleportation \cite{bouwmeester1997experimental}.
We perform the experiment by preparing X$_1$ photons in the target polarization states $\ket{\phi}$ in laboratory 1 and sending them through a 15 m single-mode fiber to laboratory 2, where they interfere with the X$_2$ photons through the BSM.
At the same time, the XX$_2$ photons undergo a polarization measurement in laboratory 2, where we collect the three-fold coincidences among their detection and X$_1$-X$_2$ coincidence events in the chosen projection of the BSM.
After the X$_1$-X$_2$ interference, the joint polarization state of X$_1$-X$_2$-XX$_2$ can be written as \cite{rota2020entanglement}:
\begin{align*}
    \ket{\Psi}=\frac{1}{2}\big( \ket{\phi^+}_{X_1,X_2}\hat{I} + \ket{\psi^+}_{X_1,X_2}\hat{\sigma}_X+\\
    -i\ket{\psi^-}_{X_1,X_2}\hat{\sigma}_Y +\ket{\phi^-}_{X_1,X_2}\hat{\sigma}_Z\big)\ket{\phi}_{XX_2}
\end{align*}
where $\hat{I}$ is the identity operator and $\hat{\sigma}_{X/Y/Z}$ are the Pauli operators.
  \begin{figure}[H]
    \centering
\includegraphics[width=\columnwidth]{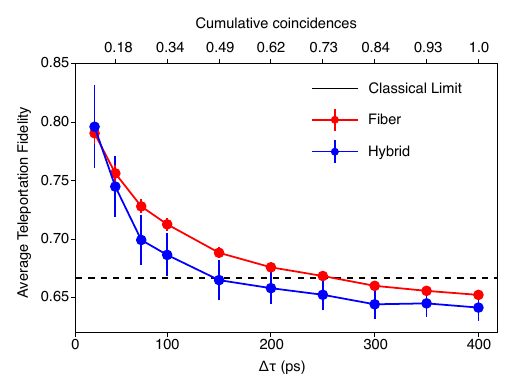}
\caption{\textbf{Temporal post-selection of quantum teleportation events.}\textit{ Teleportation fidelity averaged over the $\ket{\psi^-}$ and $\ket{\psi^+}$ projections as a function of the BSM coincidence time window $\Delta \tau$. 
We compare the results obtained for the fiber network case (red points) with the hybrid scenario (blue points), where teleported photons are sent over a 270 m free-space channel.
    On the top axis, the fraction of retained events over the total (corresponding to a coincidence time window of $400$ ps) is reported as a function of $\Delta \tau$.}}
    \label{fig:4-fidelities}
\end{figure}
The BSM projection on one of the two Bell states $\ket{\psi^-}$ and $\ket{\psi^+}$ corresponds to the application of a different operation to the teleported qubit.
The whole teleportation protocol from X$_1$ to XX$_2$ can be fully characterized by performing a quantum process tomography \cite{chuang1997prescription}, which yields the \textit{process matrix}.
By collecting the three-fold coincidences for different input states, we can reconstruct the teleportation process matrix for different values of BSM coincidence time-window $\Delta\tau$. 
Those are reported for both projections and for $\Delta \tau=30$ ps in Fig. \ref{fig:3-teleportation_results}\textbf{b}.
From these matrices, we estimate the overall teleportation fidelity which is found to be  $F_T^-=0.82\pm0.01$ for the $\ket{\psi^-}$ projection and $F_T^+=0.77\pm0.01$ for the $\ket{\psi^+}$ one.
Both values are well above the classical limit of $\frac{2}{3}$ \cite{massar1995optimal} by more than 10 standard deviations. Yet, it is worth noting that in the $\ket{\psi^+}$ projection case, we do not reach the theoretical limit of $F_T=0.827\pm0.006$ estimated for our QDs. This is due to unavoidable setup imperfections and other sources of undesired background signal.
In particular, we estimate that a slightly imperfect polarization compensation procedure may have decreased the teleportation fidelity for the $\ket{\psi^+}$ projection of around $2\%$.
We provide further discussion of these points in the SI.
  As mentioned above, our scheme uses temporal post-selection and we investigate how the teleportation fidelity changes as a function of $\Delta \tau$. The red plot in Fig. \ref{fig:4-fidelities} displays the teleportation results averaged over the $\ket{\psi^-}$ and $\ket{\psi^+}$ projections applying different coincidence time windows to the BSM up to $\Delta \tau=30$ ps. It can be clearly seen that the classical limit can be overcome by a relatively moderate temporal post-selection, with a $\Delta\tau$ of about $250$ ps. This value allows to retain more than 70$\%$ of
the total three-fold coincidence events, corresponding to $\approx30$ Hz rate of teleportation events. The best teleportation fidelity is obtained for the $\Delta \tau=30$ ps window, that allows to retain about $11\%$ of the maximum possible teleportation rate, corresponding to a $\approx4$ Hz teleportation.
These results stimulated us to explore the possibility to implement our teleportation
scheme in an urban communication scenario, specifically
using the Sapienza free-space link \cite{basso2021quantum, basset2023daylight}: a 270 m free space channel laid over the university campus in the city center of Rome.
Rather than being locally analyzed, XX$_2$ photons are sent through the quantum channel and measured in laboratory 3, in the Fermi building of Sapienza. This operation presents many challenges, including the presence of substantial losses (almost $90\%$ of the signal is lost through the channel), the need to synchronize photon detection, as well as the presence of atmospheric turbulence which makes the coupling of the signal to single mode fibers unstable. The synchronization and instability issues are alleviated by the employment of GPS disciplined oscillators and a stabilization system based on both slow and fast steering mirrors, respectively \cite{basso2021quantum}.
The results are shown in Fig. \ref{fig:4-fidelities}, where we report the teleportation fidelities we obtain for a given $\Delta \tau$, in comparison with those achieved in the fiber-only network (see SI for details). 
Given the losses present in the free-space link, the uncertainty on the fidelity values is considerably larger with respect to the fiber-only network. However, for $\Delta\tau=30$ ps, we obtain an average teleportation fidelity of $F_T = 0.80 \pm 0.04$, more than 3$\sigma$ above the classical limit. The slightly lower teleportation fidelities that can be observed in the hybrid network are related to the presence of the free-space channel, which favors the teleportation of some polarization states in comparison with others, due to imperfect polarization compensation procedures. We discuss these points in more detail in the SI.

\section*{Discussion and outlook}
In this work, we present the first demonstration of all-photonic quantum state teleportation using photons generated by independent and dissimilar quantum emitters, specifically epitaxial QDs. We also successfully scaled up the protocol in an urban quantum network that presents several challenges, including photon losses, atmospheric turbulence, and the need of synchronizing different nodes. 
We achieve this results by overcoming some of the roadblocks that have long prevented the implementation of such protocol with quantum emitters.
The key to this accomplishment was the use of strain- and magnetic-field-tunable QDs embedded in nanophotonic cavities and the exploitation of state-of-the-art techniques for single photon detection and event analysis. This allowed us to surpass the classical threshold for state teleportation while retaining more than 70$\%$ of the total events rate. Moreover, we recorded an average fidelity of up to $79 \pm1\%$ for about 11$\%$ of the total teleportation events. Our achievement now opens an avenue for the development of practical quantum networks and will stimulate additional research endeavors.
In fact, several approaches can be put in place to reach near-unity teleportation fidelities at the maximum possible rates. First, the implementation of electric fields to suppress charge noise \cite{kuhlmann2015transform} will allow improving the indistinguishably of photons generated by remote QDs. This possibility has been demonstrated for single photon sources \cite{zhai2022quantum}, but additional efforts are still needed to improve the brightness of the devices and extend the method to entangled photon sources. Regarding the free-space link, the integration of adaptive optics for compensation of high-order aberrations due to atmospheric turbulence as well as the use of efficient synchronization strategies may allow reducing photon losses to the minimum possible. Although all these tasks will require considerable effort, our demonstration of quantum teleportation in an urban communication scenario highlights that the implementation of a QD-based quantum network for quantum information processing is a likely perspective in the foreseeable future.

\printbibliography

@article{barz2012demonstration,
  title={Demonstration of blind quantum computing},
  author={Barz, Stefanie and Kashefi, Elham and Broadbent, Anne and Fitzsimons, Joseph F and Zeilinger, Anton and Walther, Philip},
  journal={Science},
  volume={335},
  number={6066},
  pages={303--308},
  year={2012},
  publisher={American Association for the Advancement of Science}
}

@article{yin2020entanglement,
  title={Entanglement-based secure quantum cryptography over 1,120 kilometres},
  author={Yin, Juan and Li, Yu-Huai and Liao, Sheng-Kai and Yang, Meng and Cao, Yuan and Zhang, Liang and Ren, Ji-Gang and Cai, Wen-Qi and Liu, Wei-Yue and Li, Shuang-Lin and others},
  journal={Nature},
  volume={582},
  number={7813},
  pages={501--505},
  year={2020},
  publisher={Nature Publishing Group UK London}
}

@article{mao2018integrating,
  title={Integrating quantum key distribution with classical communications in backbone fiber network},
  author={Mao, Yingqiu and Wang, Bi-Xiao and Zhao, Chunxu and Wang, Guangquan and Wang, Ruichun and Wang, Honghai and Zhou, Fei and Nie, Jimin and Chen, Qing and Zhao, Yong and others},
  journal={Optics express},
  volume={26},
  number={5},
  pages={6010--6020},
  year={2018},
  publisher={Optica Publishing Group}
}

@article{paraiso2021photonic,
  title={A photonic integrated quantum secure communication system},
  author={Paraiso, Taofiq K and Roger, Thomas and Marangon, Davide G and De Marco, Innocenzo and Sanzaro, Mirko and Woodward, Robert I and Dynes, James F and Yuan, Zhiliang and Shields, Andrew J},
  journal={Nature Photonics},
  volume={15},
  number={11},
  pages={850--856},
  year={2021},
  publisher={Nature Publishing Group UK London}
}

@article{schweickert2018demand,
  title={On-demand generation of background-free single photons from a solid-state source},
  author={Schweickert, Lucas and Jöns, Klaus D and Zeuner, Katharina D and Covre da Silva, Saimon Filipe and Huang, Huiying and Lettner, Thomas and Reindl, Marcus and Zichi, Julien and Trotta, Rinaldo and Rastelli, Armando and others},
  journal={Applied Physics Letters},
  volume={112},
  number={9},
  year={2018},
  publisher={AIP Publishing}
}

@article{trotta2012universal,
  title={Universal Recovery of the Energy-Level Degeneracy of Bright Excitons in InGaAs Quantum Dots without a Structure Symmetry},
  author={Trotta, R and Zallo, E and Ortix, C and Atkinson, Paola and Plumhof, JD and Van den Brink, J and Rastelli, A and Schmidt, OG},
  journal={Physical Review Letters},
  volume={109},
  number={14},
  pages={147401},
  year={2012},
  publisher={APS}
}

@article{muller2014demand,
  title={On-demand generation of indistinguishable polarization-entangled photon pairs},
  author={Glässl, M and Michler, P},
  journal={Nature Photonics},
  volume={8},
  number={3},
  pages={224--228},
  year={2014},
  publisher={Nature Publishing Group UK London}
}

@article{kimble2008quantum,
  title={The quantum internet},
  author={Kimble, H Jeff},
  journal={Nature},
  volume={453},
  number={7198},
  pages={1023--1030},
  year={2008},
  publisher={Nature Publishing Group}
}

@article{gao2013quantum,
  title={Quantum teleportation from a propagating photon to a solid-state spin qubit},
  author={Gao, WB and Fallahi, Parisa and Togan, Emre and Delteil, Aymeric and Chin, YS and Miguel-Sanchez, Javier and Imamoglu, A},
  journal={Nature Communications},
  volume={4},
  number={1},
  pages={2744},
  year={2013},
  publisher={Nature Publishing Group UK London}
}

@article{wehner2018quantum,
  title={Quantum internet: A vision for the road ahead},
  author={Wehner, Stephanie and Elkouss, David and Hanson, Ronald},
  journal={Science},
  volume={362},
  number={6412},
  pages={9288},
  year={2018},
  publisher={American Association for the Advancement of Science}
}

@article{scholl2019resonance,
  title={Resonance fluorescence of GaAs quantum dots with near-unity photon indistinguishability},
  author={Schöll, Eva and Hanschke, Lukas and Schweickert, Lucas and Zeuner, Katharina D and Reindl, Marcus and Covre da Silva, Saimon Filipe and Lettner, Thomas and Trotta, Rinaldo and Finley, Jonathan J and Müller, Kai and others},
  journal={Nano Letters},
  volume={19},
  number={4},
  pages={2404--2410},
  year={2019},
  publisher={ACS Publications}
}

@article{zhai2022quantum,
  title={Quantum interference of identical photons from remote GaAs quantum dots},
  author={Zhai, Liang and Nguyen, Giang N and Spinnler, Clemens and Ritzmann, Julian and Löbl, Matthias C and Wieck, Andreas D and Ludwig, Arne and Javadi, Alisa and Warburton, Richard J},
  journal={Nature Nanotechnology},
  volume={17},
  number={8},
  pages={829--833},
  year={2022},
  publisher={Nature Publishing Group UK London}
}

@article{maring2024versatile,
  title={A versatile single-photon-based quantum computing platform},
  author={Maring, Nicolas and Fyrillas, Andreas and Pont, Mathias and Ivanov, Edouard and Stepanov, Petr and Margaria, Nico and Hease, William and Pishchagin, Anton and Lemaitre, Aristide and Sagnes, Isabelle and others},
  journal={Nature Photonics},
  volume={18},
  number={6},
  pages={603--609},
  year={2024},
  publisher={Nature Publishing Group UK London}
}

@article{yin2017satellite,
  title={Satellite-based entanglement distribution over 1200 kilometers},
  author={Yin, Juan and Cao, Yuan and Li, Yu-Huai and Liao, Sheng-Kai and Zhang, Liang and Ren, Ji-Gang and Cai, Wen-Qi and Liu, Wei-Yue and Li, Bo and Dai, Hui and others},
  journal={Science},
  volume={356},
  number={6343},
  pages={1140--1144},
  year={2017},
  publisher={American Association for the Advancement of Science}
}

@article{dousse2010ultrabright,
  title={Ultrabright source of entangled photon pairs},
  author={Dousse, Adrien and Suffczynski, Jan and Beveratos, Alexios and Krebs, Olivier and Lemaitre, Aristide and Sagnes, Isabelle and Bloch, Jacqueline and Voisin, Paul and Senellart, Pascale},
  journal={Nature},
  volume={466},
  number={7303},
  pages={217--220},
  year={2010},
  publisher={Nature Publishing Group UK London}
}

@article{tomm2021bright,
  title={A bright and fast source of coherent single photons},
  author={Tomm, Natasha and Javadi, Alisa and Antoniadis, Nadia Olympia and Najer, Daniel and Löbl, Matthias Christian and Korsch, Alexander Rolf and Schott, Rudiger and Valentin, Sascha Rene and Wieck, Andreas Dirk and Ludwig, Arne and others},
  journal={Nature Nanotechnology},
  volume={16},
  number={4},
  pages={399--403},
  year={2021},
  publisher={Nature Publishing Group UK London}
}

@article{liu2019solid,
  title={A solid-state source of strongly entangled photon pairs with high brightness and indistinguishability},
  author={Liu, Jin and Su, Rongbin and Wei, Yuming and Yao, Beimeng and Silva, Saimon Filipe Covre da and Yu, Ying and Iles-Smith, Jake and Srinivasan, Kartik and Rastelli, Armando and Li, Juntao and others},
  journal={Nature Nanotechnology},
  volume={14},
  number={6},
  pages={586--593},
  year={2019},
  publisher={Nature Publishing Group UK London}
}

@article{gurioli2019droplet,
  title={Droplet epitaxy of semiconductor nanostructures for quantum photonic devices},
  author={Gurioli, Massimo and Wang, Zhiming and Rastelli, Armando and Kuroda, Takashi and Sanguinetti, Stefano},
  journal={Nature Materials},
  volume={18},
  number={8},
  pages={799--810},
  year={2019},
  publisher={Nature Publishing Group UK London}
}

@article{schliwa2009few,
  title={Few-particle energies versus geometry and composition of In x Ga 1-x As/GaAs self-organized quantum dots},
  author={Schliwa, Andrei and Winkelnkemper, Momme and Bimberg, Dieter},
  journal={Physical Review B},
  volume={79},
  number={7},
  pages={075443},
  year={2009},
  publisher={APS}
}

@article{legero2004quantum,
  title={Quantum beat of two single photons},
  author={Legero, Thomas and Wilk, Tatjana and Hennrich, Markus and Rempe, Gerhard and Kuhn, Axel},
  journal={Physical Review Letters},
  volume={93},
  number={7},
  pages={070503},
  year={2004},
  publisher={APS}
}

@article{scholl2020crux,
  title={Crux of using the cascaded emission of a three-level quantum ladder system to generate indistinguishable photons},
  author={Schöll, Eva and Schweickert, Lucas and Hanschke, Lukas and Zeuner, Katharina D and Sbresny, Friedrich and Lettner, Thomas and Trivedi, Rahul and Reindl, Marcus and Covre da Silva, Saimon Filipe and Trotta, Rinaldo and others},
  journal={Physical Review Letters},
  volume={125},
  number={23},
  pages={233605},
  year={2020},
  publisher={APS}
}

@article{bouwmeester1997experimental,
  title={Experimental quantum teleportation},
  author={Bouwmeester, Dik and Pan, Jian-Wei and Mattle, Klaus and Eibl, Manfred and Weinfurter, Harald and Zeilinger, Anton},
  journal={Nature},
  volume={390},
  number={6660},
  pages={575--579},
  year={1997},
  publisher={Nature Publishing Group UK London}
}

@article{rota2020entanglement,
  title={Entanglement teleportation with photons from quantum dots: toward a solid-state based quantum network},
  author={Rota, Michele B and Basset, Francesco Basso and Tedeschi, Davide and Trotta, Rinaldo},
  journal={IEEE J. Sel. Top. Quantum Electron.},
  volume={26},
  number={3},
  pages={1--16},
  year={2020},
  publisher={IEEE}
}

@article{ren2017ground,
  title={Ground-to-satellite quantum teleportation},
  author={Ren, Ji-Gang and Xu, Ping and Yong, Hai-Lin and Zhang, Liang and Liao, Sheng-Kai and Yin, Juan and Liu, Wei-Yue and Cai, Wen-Qi and Yang, Meng and Li, Li and others},
  journal={Nature},
  volume={549},
  number={7670},
  pages={70--73},
  year={2017},
  publisher={Nature Publishing Group UK London}
}

@article{massar1995optimal,
  title={Optimal extraction of information from finite quantum ensembles},
  author={Massar, Serge and Popescu, Sandu},
  journal={Physical Review Letters},
  volume={74},
  number={8},
  pages={1259},
  year={1995},
  publisher={APS}
}

@article{ates2009post,
  title={Post-Selected Indistinguishable Photons from the Resonance Fluorescence of a Single Quantum Dot in a Microcavity},
  author={Ates, S and Ulrich, SM and Reitzenstein, S and Löffler, A and Forchel, A and Michler, P},
  journal={Physical Review Letters},
  volume={103},
  number={16},
  pages={167402},
  year={2009},
  publisher={APS}
}

@article{mckeever2004deterministic,
  title={Deterministic generation of single photons from one atom trapped in a cavity},
  author={McKeever, J and Boca, A and Boozer, AD and Miller, R and Buck, JR and Kuzmich, A and Kimble, HJ},
  journal={Science},
  volume={303},
  number={5666},
  pages={1992--1994},
  year={2004},
  publisher={American Association for the Advancement of Science}
}

@article{gregersen2016broadband,
  title={A broadband tapered nanocavity for efficient nonclassical light emission},
  author={Gregersen, Niels and McCutcheon, Dara PS and Mork, Jesper and Gerard, Jean-Michel and Claudon, Julien},
  journal={Optics Express},
  volume={24},
  number={18},
  pages={20904--20924},
  year={2016},
  publisher={Optica Publishing Group}
}

@article{kuhlmann2015transform,
  title={Transform-limited single photons from a single quantum dot},
  author={Kuhlmann, Andreas V and Prechtel, Jonathan H and Houel, Julien and Ludwig, Arne and Reuter, Dirk and Wieck, Andreas D and Warburton, Richard J},
  journal={Nature Communications},
  volume={6},
  number={1},
  pages={8204},
  year={2015},
  publisher={Nature Publishing Group UK London}
}

@article{somaschi2016near,
  title={Near-optimal single-photon sources in the solid state},
  author={Somaschi, Niccolo and Giesz, Valerian and De Santis, Lorenzo and Loredo, JC and Almeida, Marcelo P and Hornecker, Gaston and Portalupi, S Luca and Grange, Thomas and Anton, Carlos and Demory, Justin and others},
  journal={Nature Photonics},
  volume={10},
  number={5},
  pages={340--345},
  year={2016},
  publisher={Nature Publishing Group UK London}
}

@article{bayer1998exciton,
  title={Exciton binding energies and diamagnetic shifts in semiconductor quantum wires and quantum dots},
  author={Bayer, M and Walck, SN and Reinecke, TL and Forchel, A},
  journal={Physical Review B},
  volume={57},
  number={11},
  pages={6584},
  year={1998},
  publisher={APS}
}

@article{schimpf2021quantum,
  title={Quantum dots as potential sources of strongly entangled photons: Perspectives and challenges for applications in quantum networks},
  author={Schimpf, Christian and Reindl, Marcus and Basso Basset, Francesco and Jöns, Klaus D and Trotta, Rinaldo and Rastelli, Armando},
  journal={Applied Physics Letters},
  volume={118},
  number={10},
  year={2021},
  publisher={AIP Publishing}
}

@article{huber2017highly,
  title={Highly indistinguishable and strongly entangled photons from symmetric GaAs quantum dots},
  author={Huber, Daniel and Reindl, Marcus and Huo, Yongheng and Huang, Huiying and Wildmann, Johannes S and Schmidt, Oliver G and Rastelli, Armando and Trotta, Rinaldo},
  journal={Nature Communications},
  volume={8},
  number={1},
  pages={15506},
  year={2017},
  publisher={Nature Publishing Group UK London}
}

@article{santori2002indistinguishable,
  title={Indistinguishable photons from a single-photon device},
  author={Santori, Charles and Fattal, David and Vuckovic, Jelena and Solomon, Glenn S and Yamamoto, Yoshihisa},
  journal={Nature},
  volume={419},
  number={6907},
  pages={594--597},
  year={2002},
  publisher={Nature Publishing Group UK London}
}

@article{gazzano2013bright,
  title={Bright solid-state sources of indistinguishable single photons},
  author={Gazzano, O and Michaelis de Vasconcellos, S and Arnold, C and Nowak, A and Galopin, E and Sagnes, I and Lanco, L and Lemaitre, A and Senellart, P},
  journal={Nature Communications},
  volume={4},
  number={1},
  pages={1425},
  year={2013},
  publisher={Nature Publishing Group UK London}
}

@article{rota2024source,
  title={A source of entangled photons based on a cavity-enhanced and strain-tuned GaAs quantum dot},
  author={Rota, Michele B and Krieger, Tobias M and Buchinger, Quirin and Beccaceci, Mattia and Neuwirth, Julia and Huet, Helio and Horova, Nikola and Lovicu, Gabriele and Ronco, Giuseppe and Covre da Silva, Saimon F and others},
  journal={Elight},
  volume={4},
  number={1},
  pages={13},
  year={2024},
  publisher={Springer}
}

@article{bayer2002fine,
  title={Fine structure of neutral and charged excitons in self-assembled In (Ga) As/(Al) GaAs quantum dots},
  author={Bayer, M and Ortner, G and Stern, O and Kuther, A and Gorbunov, AA and Forchel, A and Hawrylak, Pawel and Fafard, S and Hinzer, K and Reinecke, TL and others},
  journal={Physical Review B},
  volume={65},
  number={19},
  pages={195315},
  year={2002},
  publisher={APS}
}

@article{huber2018strain,
  title={Strain-tunable GaAs quantum dot: A nearly dephasing-free source of entangled photon pairs on demand},
  author={Huber, Daniel and Reindl, Marcus and Covre da Silva, Saimon Filipe and Schimpf, Christian and Martin-Sanchez, Javier and Huang, Huiying and Piredda, Giovanni and Edlinger, Johannes and Rastelli, Armando and Trotta, Rinaldo},
  journal={Physical Review Letters},
  volume={121},
  number={3},
  pages={033902},
  year={2018},
  publisher={APS}
}

@article{lodahl2017quantum,
  title={Quantum-dot based photonic quantum networks},
  author={Lodahl, Peter},
  journal={Quantum Science and Technology},
  volume={3},
  number={1},
  pages={013001},
  year={2017},
  publisher={IOP Publishing}
}

@article{simon2005creating,
  title={Creating single time-bin-entangled photon pairs},
  author={Simon, Christoph and Poizat, Jean-Philippe},
  journal={Physical Review Letters},
  volume={94},
  number={3},
  pages={030502},
  year={2005},
  publisher={APS}
}

@article{tanzilli2005photonic,
  title={A photonic quantum information interface},
  author={Tanzilli, Sebastien and Tittel, Wolfgang and Halder, Matthaeus and Alibart, Olivier and Baldi, Pascal and Gisin, Nicolas and Zbinden, Hugo},
  journal={Nature},
  volume={437},
  number={7055},
  pages={116--120},
  year={2005},
  publisher={Nature Publishing Group UK London}
}

@article{mair2001entanglement,
  title={Entanglement of the orbital angular momentum states of photons},
  author={Mair, Alois and Vaziri, Alipasha and Weihs, Gregor and Zeilinger, Anton},
  journal={Nature},
  volume={412},
  number={6844},
  pages={313--316},
  year={2001},
  publisher={Nature Publishing Group UK London}
}

@article{kwiat1997hyper,
  title={Hyper-entangled states},
  author={Kwiat, Paul G},
  journal={Journal of Modern Optics},
  volume={44},
  number={11-12},
  pages={2173--2184},
  year={1997},
  publisher={Taylor and Francis}
}

@article{basset2023daylight,
  title={Daylight entanglement-based quantum key distribution with a quantum dot source},
  author={Basset, F Basso and Valeri, Mauro and Neuwirth, Julia and Polino, Emanuele and Rota, Michele B and Poderini, Davide and Pardo, Claudio and Rodari, Giovanni and Roccia, Emanuele and da Silva, SF Covre and others},
  journal={Quantum Science and Technology},
  volume={8},
  number={2},
  pages={025002},
  year={2023},
  publisher={IOP Publishing}
}

@article{basso2021quantum,
  title={Quantum key distribution with entangled photons generated on demand by a quantum dot},
  author={Basso Basset, Francesco and Valeri, Mauro and Roccia, Emanuele and Muredda, Valerio and Poderini, Davide and Neuwirth, Julia and Spagnolo, Nicolò and Rota, Michele B and Carvacho, Gonzalo and Sciarrino, Fabio and others},
  journal={Science Advances},
  volume={7},
  number={12},
  pages={eabe6379},
  year={2021},
  publisher={American Association for the Advancement of Science}
}

@article{chuang1997prescription,
  title={Prescription for experimental determination of the dynamics of a quantum black box},
  author={Chuang, Isaac L and Nielsen, Michael A},
  journal={Journal of Modern Optics},
  volume={44},
  number={11-12},
  pages={2455--2467},
  year={1997},
  publisher={Taylor and Francis}
}

@article{ding2016demand,
  title={On-demand single photons with high extraction efficiency and near-unity indistinguishability from a resonantly driven quantum dot in a micropillar},
  author={Ding, Xing and He, Yu and Duan, Z-C and Gregersen, Niels and Chen, M-C and Unsleber, S and Maier, Sebastian and Schneider, Christian and Kamp, Martin and Hofling, Sven and others},
  journal={Physical review letters},
  volume={116},
  number={2},
  pages={020401},
  year={2016},
  publisher={APS}
}

@article{jayakumar2013deterministic,
  title={Deterministic photon pairs and coherent optical control of a single quantum dot},
  author={Jayakumar, Harishankar and Predojevic, Ana and Huber, Tobias and Kauten, Thomas and Solomon, Glenn S and Weihs, Gregor},
  journal={Physical review letters},
  volume={110},
  number={13},
  pages={135505},
  year={2013},
  publisher={APS}
}

@article{gisin2007quantum,
  title={Quantum communication},
  author={Gisin, Nicolas and Thew, Rob},
  journal={Nature photonics},
  volume={1},
  number={3},
  pages={165--171},
  year={2007},
  publisher={Nature Publishing Group UK London}
}
\section*{Acknowledgements}
This project has received funding from the European Union’s Horizon 2020 research and innovation program under Grant Agreement no. 899814 (Qurope) and No. 871130 (Ascent+), and from the QuantERA II program that has received funding from the European Union’s Horizon 2020 research and innovation program under Grant Agreement No 101017733 via the project QD-E-QKD and the FFG grant no. 891366. The authors also acknowledge support from MUR (Ministero dell’Università e della
Ricerca) through the PNRR MUR project PE0000023-NQSTI, the Linz Institute of Technology (LIT), the LIT Secure and Correct Systems Lab, supported by the State of Upper Austria, the European Union’s Horizon Europe research and innovation program under EPIQUE Project GA No. 101135288, and the European Commission by project QUID (Quantum Italy Deployment) funded in the Digital Europe Programme under the
grant agreement No 101091408. This work is supported by the Deutsche Forschungsgemeinschaft (German Research Foundation) through the transregional collaborative research center TRR142/3-2022 (231447078) and the European Research Council starting grant (LiNQs, 101042672).

\section*{Author contributions}
A. L., G. R., M. B., G. D. and M. B. R. performed the experiment over the fiber network under the supervision of F. B. B. and R. T.. A. L., G. R., M. B., G. D. and M. B. R.  performed the experiment over the hybrid network on the Marconi side of the free-space channel under the supervision of F. B. B. and R. T.. P. B. and A. S. participated to the experiment on the Fermi side of the free-space channel under the supervision of G. C., N. S. and F. Sc.. F. Sa., N. C. R. and K. J. contributed to the experimental setup implementation. F. Sa. and P. B. developed the data acquisition code with supervision of F. B. B., A. L., K. J., N. S., G. C. and F. Sc.. A. L., G. R., M. B. and G. D. analyzed the data. F. B. B. developed the model for the expected quantum teleportation performances. L. C. contributed to the setup and characterization of the free-space channel. 
E. S., L. H., J. N. performed preliminary experiments and studies under the supervision of K. J. and R. T..
S. F. C. d. S. and A. R.
designed and grew the QD sample.
M. B. R., T. M. K. and Q. B. designed and processed the cavity of the QD
with the supervision of R. T., A. R., T. H. L. and S. H.. S. S. processed the micro-machined piezoelectric actuator. M. A. U. C. developed the fast single-photon detectors for the experiments.
A. L. and R. T. wrote the paper with contribution from all the authors. R. T.  conceived the experiment. A. R., F. Sc., K. J. and R. T. coordinated the project.
\end{multicols}
\end{document}